# On the Ambiguities of Incompatibility in Frequentist Inference

Alessandro Rovetta

**Correspondence details**

International Committee Against the Misuse of Statistical Significance
Bovezzo (BS), Italy, 25073
Email: alessandrorovetta@icamss.com
Phone: +39 3927112808
ORCID: https://orcid.org/0000-0002-4634-279X

**Abstract**

The interpretation of the P-value and its monotone transform s=−log₂p, or S-value, remains debated despite decades of dedicated literature. Within the neo-Fisherian framework, these values are often described as indices of (in)compatibility between the observed data and a set of ideal assumptions (i.e., the statistical model). In this regard, this paper proposes the distinction between two domains: the *model domain*, where assumptions are taken as perfectly true and every admissible outcome is, by construction, fully compatible with the model; and the *real domain*, where assumptions may fail and face empirical scrutiny. I argue that, although interpreted through an objective numerical index, any level of incompatibility can arise only in the latter domain, where the epistemic status of the model under examination is uncertain and a genuine conflict between data and hypotheses can therefore occur. The extent to which P- and S-values are taken as indicating incompatibility is a matter of contextual judgment. Within this framework, descriptive approaches serve to quantify the numerical values of P and S; these can be interpreted as indicative of a certain *degree* (or *amount*) of incompatibility between data and hypotheses once causal knowledge of the data-generating process and information about the costs and benefits of related decisions become clearer. Although the distinction between the model domain and the real domain may appear merely theoretical or even philosophical, I argue that this perspective is useful for developing a clear mental representation of how statistical estimates should be evaluated in practical settings and applications.

**Introduction**

Within the neo-Fisherian framework, a P-value is often interpreted as a descriptive index of the compatibility between observed data and a statistical model defined by a set of foundational assumptions and possibly a target assumption—or target hypothesis—about the parameter [1]. Its monotone transform $-\log_2 p$, or S-value, is interpreted instead as the quantity of information in 'bits' provided by the data against the model. This paper discusses the distinction between the formal, internal interpretation of these measures within the model domain and their operational use in the real domain, where assumptions may fail and hypotheses are subject to empirical scrutiny. In particular, it will be argued that incompatibility cannot arise within the model domain—where all admissible outcomes are, by definition, consistent with the model—but only in the real domain, as an interpretative judgment made when models are confronted with empirical reality.

**Compatibility and incompatibility**

The notion of compatibility between data and models is not new in statistics, as it can be traced back at least to Karl Pearson's 1900 work on the chi-squared test [2]. In that framework, the discrepancy between observed and expected frequencies was described as a measure of compatibility between them. Subsequent synonyms of compatibility include consistency, consonance, agreement, and coherence [1]. In recent years, a growing body of methodological work has advocated replacing the traditional language of "statistical significance" with notions of compatibility and incompatibility [3,4,5,6,7,8,9], although illuminating articles had already been published in major journals by authors of Rothman's caliber, about half a century ago [10]. The motivation is commendable: to avoid the well-documented misinterpretations of P-values as posterior probabilities or as dichotomous verdicts of "significance" or truth. By speaking of a hypothesis as more compatible or less compatible with data, the aim is to communicate the graded, descriptive nature of frequentist evidence [9]. Indeed, compatibility expresses how well data agree with a model, without implying any confidence or support for that model—which would instead impose constraints on the alternatives consistent with the observed data [1,3].

While I strongly endorse this goal, I argue that there is a conceptual tension that deserves attention about *incompatibility*. Properly understood, incompatibility refers to 'the condition of two things being so different in nature as to be incapable of coexisting.' By extension, one might also speak of degrees of incompatibility, understood not as logical impossibility but as increasing difficulty of coexistence: the greater the discrepancy, the more strained the coexistence appears. Yet we assume a statistical model M to be true, where such M is generally made of a set of foundational assumptions A (e.g., homoscedasticity and linearity) and a target hypothesis H (e.g., the null hypothesis of no effect/association/difference) [11,12]. Then, in the ideal world induced by M, a realization Y=y must be perfectly compatible with M.[1] This realization may deviate "substantially" from an expected value for Y; however, it cannot be incompatible with M in any strict logical sense.

---

[1] Provided that y $\in$ support(M).

This distinction becomes clearer if we separate two domains: the *model domain* and the *real domain*.

*The model domain*

In the model domain, the foundational assumptions A and the target hypothesis H are 100% true. Accordingly, however distant y may be from the defined Y-expectation, it remains entirely consistent with M, since it is generated under the very constraints imposed by M. Hence, there can be no incompatibility between y and M: they always coexist by construction. P-values and S-values computed under M are thus abstract, objective numerical functions of the data, reflecting the long-run frequency with which discrepancies as or more extreme than the one observed would occur under the considered model.

*The real domain*

The real domain is the empirical context in which models are applied. Here, matters differ: indeed, A may or may not be correct, and H is under scrutiny. We can therefore speak of *refutational evidence* as the degree to which the observation challenges H under A [1,13]. When the P-value is "small" or the S-value is "large," we say that the target hypothesis appears less compatible with the data under the other assumptions. Thus, we use the calculations from M, like P-values, S-values, and interval estimates, as tools to inform real-world evaluations; we do this by integrating them with other non-statistical evidence (e.g., biological plausibility) across a spectrum of choices that extends from sequential updates to terminal actions [12,14,15]. The consequential nature of these choices (e.g., the risks they entail for stakeholders) must be calibrated against the strength of the overall body of evidence; accordingly, the mere compatibility of a result with a hypothesis should never serve as the sole basis for making any decision.[2]

**A more formal definition**

Drawing on Greenland [1], the observed y can be generally treated as a data vector, realization of a random vector Y whose possible values span a sample space. Assumptions A define a family $F_A$ of probability mass functions (denoted by f) on that space, representing all distributions *consistent* with A. Here, H would represent a further constraint on $F_A$, thereby yielding the more restrictive family $F_M$ and forming a model M=(A,H) nested within A. Within this framework, I argue that the model itself never declares a data point as 'having a degree of incompatibility' with the Y-expectations under $F_A$ or $F_M$; what it provides is a measure of separation, the distance $d(y;\{\mu_f\})$, between the observed data and the set $\{\mu_f\}$ of Y-expectations, with f in $F_A$ or $F_M$.

Indeed, an observed upper-tail P-value is simply the ordinal location of the above separation statistics within the reference distribution derived from A (or M); that is, the quantile at which the observed 'd' falls. This distance 'd' is inversely related to the P-value, where the latter quantifies the probability of observing a distance D≥d under repeated sampling from the model.[3] The S-value is then the logarithmic transform of that P-value ($s=-\log_2 p$), providing a more equal-interval measure of the probabilistic information in 'bits' (each additional 'bit' halves p): the larger the value of S, the less frequent it is to observe D≥d under the model [1].

---

[2] To understand why, it is enough to note that, without integrating obvious causal information, the strong association between 'drowning deaths by falling into a pool' and 'release of films starring Nicolas Cage' is compatible with the hypothesis of his responsibility [18].

Instead, incompatibility as indicating *conflict* between data and model only arises due to an external interpretation of these numerical quantities, and *not* as a quantity assigned by the model itself. As a matter of fact, the model must be internally *consistent*—a synonym for compatibility that refers to the idea that every element generated through the model, like y, must be <u>perfectly</u> compatible with the model itself. It is we, the analysts, who interpret a "low" frequency (a "large" distance) under A (or M) as indicating some degree of conflict, as we don't know if A (or M) is true or not in practice. In this sense, incompatibility is not a property of the model–data relation, but a practical, human evaluation of that abstract, objective relation.

**Why this matters**

The problem is not trivial semantics. How we label P-values or S-values has direct consequences for decision making, because their interpretation inevitably interacts with judgments about costs and benefits concerning our choices [14,15]. A numerical value such as s = 4.3 bits is an objective calculation within the model; yet deciding whether this represents "high" or "low" incompatibility requires contextual evaluation [12,14,19,15]. In essence, this evaluation can be viewed as a continuous extension of the Neyman–Pearson choice of α, where α denotes the theoretical long-run bound for Type I errors [16,17]; here, however, instead of relying on a fixed threshold, one interprets a gradation of refutational evidence that more faithfully captures the nuances of real-world evaluations [14,15].

For example, suppose I am developing a drug and obtain a point estimate of 0.63 and a 95% interval estimate for the risk ratio (RR) of a severe side effect as (0.40, 1.00). Given the foundational assumptions A, this 95% interval represents the range of hypothetical parameters under which the observed data would be roughly less "surprising" than 4 heads in 4 coin tosses would be under the assumption 'no bias towards heads' [14,19]. If I consider the hypothesis of a detrimental effect RR = 1.20 under A, I thus know that the data provide more than 4 bits of refutational information against it, as 1.20 lies outside the 95% interval. But is this degree of incompatibility sufficient to consider the hypothesis implausible in practice? The answer depends on how costly it would be to proceed under a false sense of safety versus how costly it would be to halt development prematurely [14,15,19]. It further hinges on our confidence in A and on broader considerations that cannot be settled by statistics alone [12,14,20].

This illustrates that operational incompatibility is not merely a statistical construct but also a small but essential building block in bridging analysis to decision-making [14,15]. The transition from a numerical index of distance to a qualitative judgment of evidential weight cannot be separated from the researcher's evaluation of costs and benefits and the causal context [14,20].

---

[3] When M adds to A an interval restriction $[x_1,x_2]$ on a hypothesis about the parameter μ, we have $d([x_1,x_2];\mu) = \min \{d(x;\mu) : x \in [x_1,x_2]\}$.

In this regard, we can distinguish between two notions of compatibility. The first could be called *logical* compatibility: any y that is supposed to be generated under the model must necessarily be perfectly compatible with the model itself. As said, this is a requirement of logical internal consistency. P-values and S-values are then objective functions of the data under that model, quantifying numerical properties such as frequency and discrepancy (often in units of standard errors, as in Wald-type analyses) rather than logical conflict. The second notion of compatibility could be called *operational* compatibility: within the real domain, analysts interpret "small" P-values or "large" S-values as statistical evidence of reduced compatibility or increased refutational evidence against the model (or against the target hypothesis under the other assumptions), with the weight attributed to this information depending on the estimated costs and benefits as well as the knowledge of the physical process that generates the data (data-generator).[4]

**There is still room and need for a descriptive approach**

In practice, researchers are often confronted with layers of uncertainty that include—and extend well beyond—the balance of costs and benefits [14,20]. This happens even under the most ideal conditions, including randomization [21,22,23]. The very set of assumptions A that underpin a statistical model, which should be ideally derived from the data-generator, is often highly uncertain; so much so that it is not uncommon to find multiple defensible models yielding markedly different results [19,20,24,25]. Accordingly, analysts should generally not proceed to definitive judgments based on single studies or limited evidence [26]. Therefore, a descriptive stance remains valuable and necessary [1,9,19,27].

By quantifying the refutational information implied by the data under a given model, one can document how observations diverge from a set of hypotheses without claiming more than the model permits. This function goes beyond the reporting of a discrepancy, since it provides a calibrated evidential scale while permitting that future analyses, ideally grounded in better causal knowledge and clearer cost–benefit specifications, may better interpret the same results [14,27]. Thus, descriptive and evaluative roles should not be seen as competing but as complementary: the former secures transparency across models and studies, while the latter becomes viable only when the surrounding decision and modeling context is sufficiently well defined [19].

---

[4] Note that, within the support of the model, just as we cannot confirm the model on the basis of the sole P-value (since even p=1 does not indicate support but only perfect data–model coherence), it is equally impossible to epistemically reject it on the basis of the sole S-value. Indeed, perfect incompatibility (or impossibility) would require s=+∞. Therefore, the choice to regard a model as 'sufficiently incompatible' with the data must be motivated by more complex, situational considerations.

Certainly, since the very decision to continue or discontinue a project is partly based on the observed results, this approach may be described as *quasi*-unconditional, since it aims to restrict human influence on analysis and subsequent decisions to the strict minimum [[14]].

**Human uncertainty**

Statistical analysis is never conducted in a vacuum: it is shaped by human choices, values, and limitations. Mathematics provides results that are valid only if all assumptions hold, yet many of these assumptions are often fragile or even unrealistic. Researchers bring with them biases, incentives, and cultural habits or even rituals that heavily affect how data are generated, analyzed, and reported [12,28,29,30,31,32]. This "human factor" implies that concepts like objectivity and subjectivity, often invoked in statistical debates, are better understood as bundles of more precise attributes such as transparency, fairness, and context awareness [28,29].

Therefore, being tied to the real domain, any incompatibility assessment is subject to both scientific uncertainty and the web of human uncertainties linked to cognitive, emotional, social, and economic factors. While the numerical value of P or S, which we understand as quantifying compatibility or refutational evidence, is always exact and undisputable (under the assumption of no computational errors), the assessment of its magnitude and its practical weight is—and must be—a matter for informed discussion. This further underscores the caution advocated by proponents of descriptive, unconditional approaches [9,14,19,27,15,34].

**Data and results are always true, yet inherently uncertain**

Data are objective records of whatever process generated them; any "falsity" concerns our description and use of those records, not the records themselves. Even when shaped by questionable practices such as P-hacking—so prevalent in the scientific literature [33]—data remain valuable, meaning they still carry information about the mechanisms that physically produced them [20]. In this sense, there are no inherently "false" data, only misguided interpretations or misleading presentations. Given the many unknowns about the steps that led to any given observation or result, the common expression "the data speak for themselves" represents a dangerous mystification: it portrays data as the reflection of an objective, universal procedure that leaves no room for doubt or debate. By contrast, it is essential to recognize—and to embed in routine practice—that even claims of mere support for a hypothesis require far more than data–model compatibility; at a minimum, they demand clear information about the data-generator.

The stakes here extend beyond ethics: they are deeply epidemiological, with concrete consequences for public health and the credibility of science itself [26,27,28,33,35,36]. Misinterpretation or misuse of data can lead to harmful policies, undermining health outcomes and eroding public trust in scientific institutions. These aspects cannot be ignored when drawing inferences from data—whether statistical or otherwise—and they demand ongoing scrutiny and investigation, as well as an attitude of humility, transparency, collaboration, and openness to revision [9,14,19,27,15,34]. This strongly underscores that narrowly defined issues such as statistical incompatibility, though rightly situated in the real domain, are only small pieces in the much larger puzzle of building scientific evidence.

**The model is perfect, but our world is not**

The distinction between objective measures such as 'frequency' and 'distance' and evaluative notions such as 'incompatibility' and 'refutational evidence' is intended to make more explicit the net separation between the abstract, ideal world assumed by the statistical methods we employ and the practical, uncertain reality we inhabit. This distinction is particularly relevant in light of the long history of misinterpretations and abuses of frequentist statistics due to the persistent, wishful search for objective methods to discern "true" from "false," a search that too often collapses the nuances of a continuous reality into dichotomous judgments of "yes" and "no" [30,36]. The aim is to provide a clear formal representation of the distinction, as a reminder of two consequential aspects. The first is the profound level of idealization that characterizes the theoretical data-generator and thus the interpretability of quantities like P-values and S-values [20]. The second is the extent to which the assumptions underlying those measures must be strong in practice to be used to inform highly consequential decisions: enough to overcome overall epistemic uncertainty and approach the supposed level of idealization [12,14,37].

**About the relationship between S-values and surprise**

S-values are also referred to as 'surprisal' and have been described as measures of divergence, incompatibility, conflict, refutational evidence or information, and 'surprise' [1,14,19,38,37]. Even where the term 'surprise' is not explicitly mentioned, formulations such as 'the result is as *surprising* as observing *s* heads in *s* fair coin tosses' establish a direct link to the notion of surprise, both semantically and conceptually. However, as argued here, all these terms carry nuances—or even substantially different meanings—because they are embedded in distinct epistemic contexts.

Divergence, understood as a 'distance' in a model-defined space, is an objective measure insofar as it is abstract. By contrast, the remaining terms refer to interpretations made in the real domain based on the value of S calculated in an ideal world (that induced by the model, under the observed data). The concept of 'surprise' may even add an additional layer of complexity, as it refers to a human sensation with strong degrees of subjectivity. This could expose the term to the same risks of mystification that have affected "statistical significance," which has often been mistaken for practical importance.

Nonetheless, I also underscore that surprisals incorporate objective components, and that the prominence given to the subjective dimensions in evaluating surprise can confer operational benefits. First, the concept of surprise within the surprisal framework is not vacuous, but arises from comparison with a concrete and cognitively simple physical experiment: the coin toss. Surprisal does not quantify surprise on a universal scale, with gradations establishing whether the number 's' represents a "high" or "low" degree of surprise; rather, it provides a comparison, a benchmark against something we know well and have direct perceptual experience with. The requirement of internal consistency reduces to the modest condition that 's+1' represents a result *more* surprising than 's' when compared with the model's prediction.

Secondly, the notion of surprisal underscores that our evaluation of surprise—and thus of the risks we assume when making a "scientific bet"—is highly situational; more formally, that loss acceptability, much like the choice of an endpoint in medicine, inevitably entails margins of subjectivity [14,15,39]. For instance, how many consecutive heads 's' would I need to observe in as many tosses before starting to suspect that such tosses are biased towards heads? The answer depends both on the stakes and on one's cognitive stance. If it is merely an epistemic curiosity, I might settle for 4 or 5 heads in a row; but if the wager involves a high cost for erring in the direction of false positives (e.g., concluding that the coin tosses are biased when they are not), then I might adopt a more conservative approach, such as requiring 7 or 8 consecutive heads in as many tosses.[5] The key point is that such responses are expected to vary substantially across—and even within—individuals, as they are conditioned by evaluations and perceptions of costs and risks that are shaped by the specific psycho-social and cultural context [40].

**Notes on compatibility intervals as surprisal intervals**

Compatibility intervals refer to the traditional $(1-\alpha)100\%$ 'confidence' intervals, understood as the range of hypotheses with $p>\alpha$ under the data and model [1,5,10,19,41,42]. Compatibility and incompatibility are bidirectional properties. If the observed data have a certain degree of (in)compatibility with the hypothesis, then it is equally true that the hypothesis has a certain degree of incompatibility with the data. For example, a 95% compatibility interval can be described as 'the range of hypotheses least incompatible with the data under the $p>0.05$ criterion and the model.' Given the earlier definitions of the S-value as a measure of surprise and incompatibility [1,38], this opens the door to expressions such as 'the range of hypotheses least surprising compared with the data under the $s<4.3$ bits criterion (since $-\log_2 0.05 \approx 4.3$) and the model' [13,14].

In this regard, it is important to specify that such a definition in no way implies a posterior probability for the hypotheses: the statement refers to the real domain, where the epistemic state of the hypothesis about the parameter is uncertain. This leads us to assess the degree of incompatibility of such hypothesis with the observed data through an ideal model in which that hypothesis is true. Therefore, in the real domain, the hypothesis under examination acts as a *generator of surprise* (and thus is *surprising* to some degree) under the data and the underlying assumptions [13].

This can be better appreciated by analyzing the physical process of inspecting an interval estimate: formally, an interval estimate does not merely display a range of hypotheses about the parameter, but rather a range of models that incorporate a given hypothesis about the parameter. Clearly, since each hypothesis about the parameter is different, these models cannot all be true at the same time. It is we, the analysts, who examine how incompatibility (and hence surprise, as quantified by surprisal) would change as a function of different models. In the real domain, each model within the interval estimate acts as a generator of surprise with respect to the observed data (which are fixed before the interval inspection process). No assessment of the probability or validity of such models under the data has been made, and it would be a serious error to think otherwise. We are merely acknowledging a degree of conflict between the model and the data in the real domain, where the numerical index we use to quantify the "real" conflict is determined by the model in its ideal domain, under the observed data.

---

[5] Note that I have exploited a false dichotomy here: the issue is not merely whether the tosses are biased or not, but also how strongly they are biased. This relates to the evaluation of how important the discrepancy is between the observed and the expected number of heads.

**Conclusion**

P-values and S-values provide abstract, objective information about the frequency with which results as or more extreme than the one observed would occur under a given set of ideal assumptions. This information can be treated as a measure of (in)compatibility between the observed data and the considered model. However, incompatibility reflects an external judgment in a real domain, since every outcome ideally generated according to the rules imposed by the model is, by construction, entirely compatible with such model. In empirical practice, the epistemic status of the model is instead uncertain, and thus P- and S-values are employed as mere components within a broader investigative framework (e.g., a framework that aims to assess the model's adequacy for the research purpose). While the values of P and S are objective insofar as they are abstract functions of the data under an ideal perfect model, the practical weight accorded to the degree of (in)compatibility they convey is contingent on context-dependent appraisals; the latter should evolve as information about costs and benefits becomes clearer and on the extent to which the working model provides a sufficiently good approximation to the ideal one. Such process of refinement can be slow in sciences characterized by high variability and situational complexity, emphasizing the need for purely descriptive approaches that document the amount of refutational information (e.g., the number of 'bits') provided by the data against the model without attaching any immediate practical weight. The ultimate goal is to create a durable evidential record that can later be integrated with causal knowledge to inform sound decision-making.

**Conflicts of interest**

The author declares no conflicts of interest.

**Acknowledgments**